Brownian simulations for fracture of star polymer phantom networks

Yuichi Masubuchi*, Yusuke Koide, Takato Ishida, and Takashi Uneyama,
Department of Materials Physics, Nagoya University, Nagoya 4648603, JAPAN

* To whom correspondence should be addressed.

Ver. Sep 19, 2024

**Abstract**

Based on a recent simulation study [Masubuchi et al., Macromolecules, 56, 9359 (2023)], the cycle rank plays a significant role in determining the fracture characteristics of network polymers. However, the study only considered energy-minimized networks without the effects of thermal agitation. We conducted Brownian dynamics simulations at various stretch rates to address this gap. The results showed that even with Brownian motion, the strain and stress at the break obtained for different node functionalities and conversion ratios exhibited master curves if plotted against cycle rank. These master curves were dependent on the strain rate, with the curves tending to approach those observed in energy-minimized simulations as the strain rate decreased, even though the fracture process was affected by the competition against Brownian motion, elongation, and bond degradation.

**Introduction**

Numerous studies have sought to explore the fracture and rupture of network polymers through molecular simulations. The majority of these studies focus on inter and intra-chain interactions. Stevens [1,2] conducted coarse-grained bead-spring simulations with excluded volume to rupture densely cross-linked polymer networks. Tsige et al. [3,4] expanded on this method to investigate the effects of the cross-linker functionality. Dirama et al.[5] introduced ionic interactions, while Mukherji and Abrams[6] incorporated bending rigidity into their simulations. Yang and Qu[7] also explored bending rigidity. Sliozberg et al. [8] studied the effects of trapped entanglements, echoing the work of Solar et al. [9], who performed similar simulations. Following Stevens[1], Lin et al. [10] discussed the impact of the shortest path via bead-spring simulations. Arora et al.[11] and Barney et al.[12] examined the significance of loops. Wang et al.[13] focused on dangling ends. Atomistic modeling has been utilized for epoxy resins in studies by Nouri[14] and Moller[15]. There have also been investigations on phantom chains, with Masubuchi et al.[16] studying the effect of molecular weight distribution of network strands on rupture.

The studies mentioned above consider the Brownian motion of polymers, regardless of differences in intra and inter-molecular interactions. The strain rate influences the results, as explicitly discussed in some studies[11]. Even at low strain rates where entropic elasticity is reasonably attained for each unbroken strand, the stress includes contributions from structural relaxation caused by each bond scission. When network percolation is eliminated, the relaxation time is similar to that of a polymer with the molecular weight of the entire simulated system. Consequently, the elimination of strain rate dependence is practically challenging.

Some studies[17–20] employed simulation methods extracting the force balance network without Brownian motion to avoid difficulties due to the dynamics. Apart from the studies focusing on modulus[17–19], some results have been reported on network rupture[20–26]. For instance, Masubuchi et al.[21] employed such a method for networks made from star-branched prepolymers with various branching functionalities and conversion ratios. They found that the fracture characteristics, which include strain and stress at break and work for breakage of the examined networks, can be summarized as master curves if plotted against the cycle rank of the networks. Although this finding on the significance of cycle rank is intriguing, the results may depend on the model and the stretching scheme, as the fracture behavior reflects molecular interactions and microscopic dynamics.

This study focuses on the relations between fracture characteristics and cycle rank for the case with Brownian motion, which was incorporated into the phantom chain network model used in previous studies[20–25]. The network rupture was observed under various strain rates, and the stress-strain relation was recorded. The stress and strain at the break were defined from the peak of the obtained curve. For the fracture characteristics, the relationship between cycle rank was discussed. The results demonstrate that the fracture characteristics draw master curves as functions of cycle rank even with Brownian motion, although depending on the strain rate. Details are shown below.

**Model and Simulations**

The networks examined in this study were created from Rouse-Ham-type star branch prepolymers with the number of branching arms $f$ and the number of beads per arm $N_a$. Sols of star polymers were placed in a simulation box with periodic boundary conditions and the bead number density $\rho$. The position of each bead $\mathbf{R}_i$ obeyed the Langevin equation of motion written below.

$$0 = -\zeta(\dot{\mathbf{R}}_i - \boldsymbol{\kappa}\mathbf{R}_i) + \frac{3k_BT}{a^2}\sum_k f_{ik}\mathbf{b}_{ik} + \mathbf{F}_i \qquad (1)$$

Here, $\zeta$ is the friction coefficient, $\boldsymbol{\kappa}$ is the velocity gradient tensor, $k_BT$ is thermal energy, and $a$ is the average bond length. $\mathbf{b}_{ik}$ is the bond vector defined as $\mathbf{b}_{ik} \equiv \mathbf{R}_i - \mathbf{R}_k$, and $f_{ik}$ is the non-linear spring coefficient written as $f_{ik} = (1 - \mathbf{b}_{ik}^2/b_{\max}^2)^{-1}$ with the maximum stretch $b_{\max}$. This nonlinearity is necessary to suppress bond breakage by thermal agitation. $\mathbf{F}_i$ is the Gaussian random force obeying $\langle \mathbf{F}_i \rangle = \mathbf{0}$ and $\langle \mathbf{F}_i(t)\mathbf{F}_j(t') \rangle = 2k_B T \delta_{ij} \delta(t-t') \mathbf{I}/\zeta$ with the unit tensor $\mathbf{I}$. From the parameters defined above, length, energy, and time units are chosen as the average bond length $a$, thermal energy $k_B T$, and diffusion time of the single bead $\tau = \zeta a^2/k_B T$. Quantities hereafter are normalized according to these units.

End-linking reactions were introduced for the equilibrated sols, as performed previously[16,27]. According to the experimental strategy taken for tetra-PEG gels[28,29], half of the star polymers were colored to eliminate the formation of primary loops, and the reaction took place only between prepolymers with colored and uncolored polymers. During the Brownian dynamics simulation, when a pair of chain ends came closer than a reaction distance $r_r$, the examined chain ends were connected by a spring with the reaction probability $p_r$. The spring constant for this newly introduced connectivity is the same as that for the chain connectivity. The snapshots were taken at various conversion rates $\varphi_c$ that is defined as the ratio of reacted chain ends to all the reactive segments in the system. For the networks thus created, the uniaxial stretch was imposed with the strain rate $\dot{\varepsilon}$. During elongation, bonds were removed when the bond length exceeded a critical value $b_c$.

For comparison, simulations with energy-minimization were also conducted. The minimized total energy is consistent with the non-linear spring, as shown below.

$$U = -\frac{3k_B T b_{\max}^2}{2a^2} \sum_{i,k} \ln\left(1 - \frac{\mathbf{b}_{ik}^2}{b_{\max}^2}\right) \qquad (2)$$

The employed minimization scheme is the Broyden-Fletcher-Goldfarb-Sanno method[30], with the bead replacement parameter $\Delta r$ and the energy conversion parameter $\Delta u$.

The parameters were chosen as follows: $f$ was varied between 3 and 8, and $N_a$ was fixed at 5. The system size was determined to include 1680 branching arms irrespective of $f$. The other parameters were $\rho = 8$, $b_{max} = 2$, $r_r = 1$, $p_r = 0.1$, and $b_c = \sqrt{3.6}$. A 2nd-order scheme[31] was employed for the numerical integration of eq. 1. The step size $\Delta t$ was chosen at 0.01 for the equilibration and gelation stages and was reduced to 0.001 during elongation to accommodate the spring non-linearity. For energy-minimized simulations, the parameters were chosen at $\Delta r = 0.01$, and $\Delta u = 10^{-4}$. The non-linear spring constant was truncated at $f_{ik} = 10$ for numerical convenience, and the value is consistent with the choice of $b_c$.

**Results and Discussion**

Figure 1 exhibits typical behavior for the network with $f = 4$ and $\varphi_c = 0.95$, stretched under the strain rate at $\dot{\varepsilon} = 10^{-4}$. The stress shown in panel (a) increases as the system is stretched, as seen in panels (c)-(e), with small bumps reflecting bond breakage shown in panel (b). Then, it steeply decreases according to the elimination of percolation around $\varepsilon \sim 2.5$. The stress is not mitigated to zero due to the drag force of the remaining cluster in panel (f). Instead, an increase in stress due to the viscoelastic response is observed in the later stage in $\varepsilon > 3$. Before the breakage, no strong inhomogeneity was observed, which differed from the previously reported energy-minimized cases [20–26].

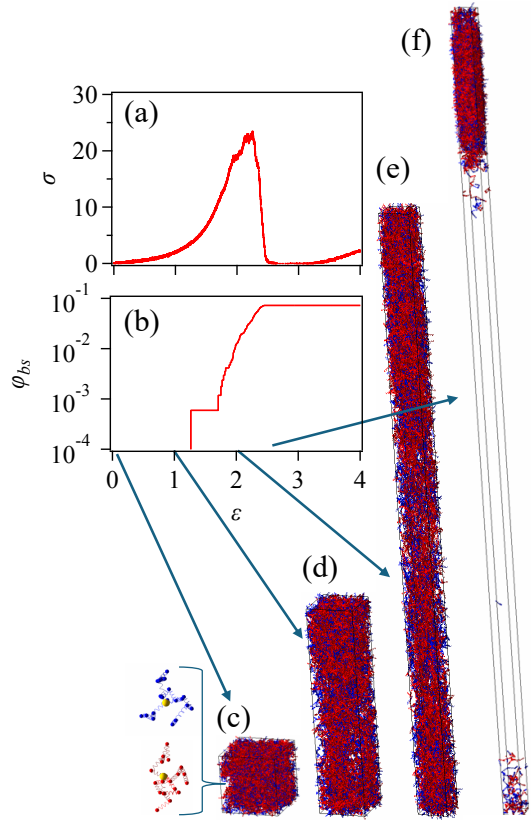

**Figure 1** Typical example of the simulation for $f = 4$, $\varphi_c = 0.95$, and $\dot{\varepsilon} = 10^{-4}$. Panels (a) and (b) exhibit the development of true stress $\sigma$ and the ratio of broken strands $\varphi_{bs}$ against true strain $\varepsilon$. Panels (c) to (f) show snapshots during the stretch at $\varepsilon = 0, 1, 2,$ and 2.5, respectively. In panel (c), a pair of prepolymers is also displayed.

Figure 2 (a) shows the stress-strain curves for a network with $f = 4$ and $\varphi_c = 0.95$ at various strain rates. The initial slope is the same in cases with Brownian motion and larger than the energy-minimized case shown by black broken curves because thermal agitation increases the average bond length. The peak stress and strain at the peak decrease with lower strain rates. Although they become apparently similar to those for the energy-minimized case at the slowest elongation, the fracture behavior is qualitatively different, as discussed below. The rate-dependent part of the stress is generated by insufficient stress propagation; the elongation develops faster than the structural reorganization due to force balance. Besides, due to the drag force term in eq. 1, stress does not decrease to zero, even after the percolation is eliminated and the stretch rate is small. See green and red curves in the region $\varepsilon > 3$. The stress after the peak is influenced by the structural relaxation of residual domains, and this influence becomes less significant as the strain rate decreases.

As a crude estimate of the structural relaxation time, let us calculate the relaxation time of a cluster of prepolymers with an end-to-end distance similar to the simulation box dimension before the stretch. The number of prepolymers in such a cluster is given by $N' = L_0^2/[(2N_a + 1)a^2]$, and the Rouse time is $\tau_R = [N'(2N_a + 1)]^2/(3\pi^2) = L_0^4/(3\pi^2) \sim 10^3$, with $L_0 \sim 12$. The result in Fig 2 (a) is consistent with this rough estimation where the stress does not mitigate to zero when $\dot{\varepsilon} > 1/\tau_R \sim 10^{-3}$. One may obtain the exact relaxation time from the connectivity matrix[32]. Nevertheless, if the peak stress and the strain at the peak are regarded as those at the break, the values decrease as the strain rate decreases, yet they differ from those for energy-minimized networks.

Figure 2 (b) exhibits the development of the broken strand ratio $\varphi_{bs}$. Bond scission occurs when the strain rate exceeds the relaxation rate of the remaining domains. Consequently, $\varphi_{bs}$ continually increases under strain rates greater than 10⁻³ while stabilizing at a specific value for lower strain rates. The final values of $\varphi_{bs}$ depend on the strain rate and differ from that observed for the energy-minimized simulations (black broken curves) even though each simulation was started from the same initial configuration. This result demonstrates that the broken network structure is affected by the kinetic pathway that differs from the energy-minimized case when thermal agitation exists. For instance, thermally induced bond breakage occurs in the Brownian simulations depending on the choice of $b_c$ and $\dot{\varepsilon}$. Namely, when the thermal degradation rate determined by $b_c$ is larger than $\dot{\varepsilon}$, $\varphi_{bs}$ starts to grow at a smaller strain. $\varphi_{bs}$ for $\dot{\varepsilon} = 10^{-4}$ (red) shows such a behavior, and it is larger than the other cases shown in green and blue.

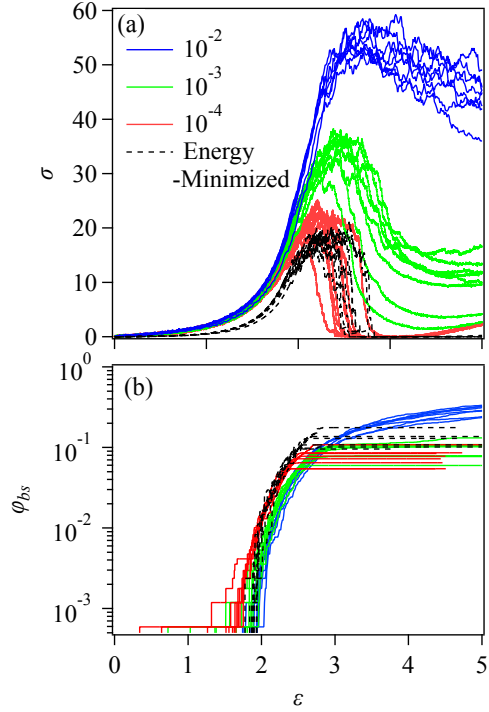

**Figure 2** Development of true stress $\sigma$ (a) and broken strand fraction $\varphi_{bs}$ (b) against true strain $\varepsilon$ for the networks with $f = 4$ and $\varphi_c = 0.95$ under elongation at the strain rates of $\dot{\varepsilon} = 10^{-4}$ (red), $10^{-3}$ (green), and $10^{-2}$ (blue), respectively. For comparison, the results from energy-minimized simulations are also shown by black broken curves.

From the stress-strain curves, the strain and stress at break, $\varepsilon_b$ and $\sigma_b$, were extracted. Since stress does not mitigate to zero, as discussed above, breakage is defined by the maximum stress. Figure 3 shows $\varepsilon_b$ and $\sigma_b$ for the cases with $f=4$ (circle) and 8 (triangle) at $\varphi_c=0.6$ (left column) and 0.95 (right column). As discussed in Figure 2 (a), $\varepsilon_b$ and $\sigma_b$ decreases with decreasing $\dot{\varepsilon}$. Note that at the smallest $\dot{\varepsilon}$, $\varepsilon_b$ and $\sigma_b$ become smaller than those obtained for the energy-minimized calculations shown by the leftmost symbols for some cases due to thermal degradation discussed above. Concerning $\varepsilon_b$, in the case with $\varphi_c=0.6$, the error bars are much larger than the other cases, reflecting intense bumps with similar peak values in the stress-strain curves. Similar behaviors were observed for the networks with small $f$ and $\varphi_c$ values. The stress fluctuation for these cases reflects the structures where the percolation is attained by sparse connectivity.

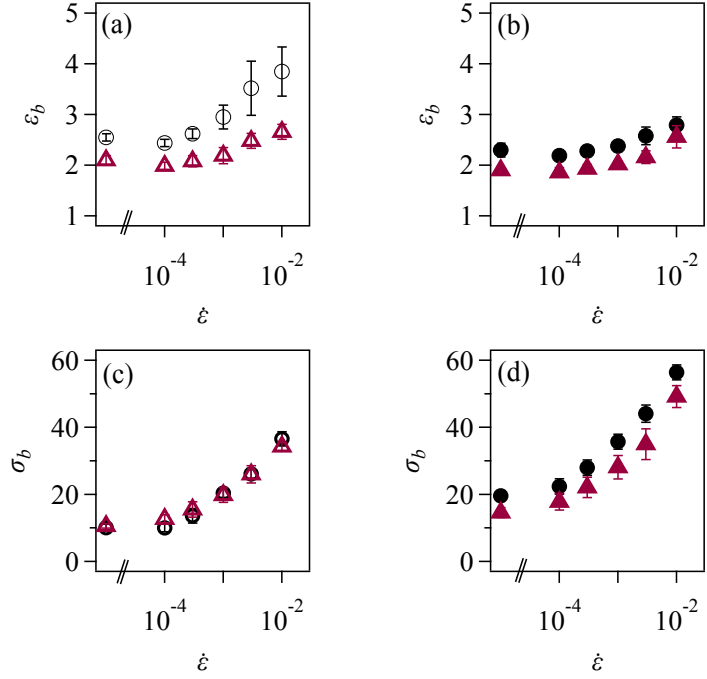

**Figure 3** Strain and stress at break, $\varepsilon_b$ (upper panels) and $\sigma_b$ (lower panels) for the networks with $f$=4 (circle) and 8 (triangle) at $\varphi_c$=0.6 (left column) and 0.95 (right column) plotted against strain rate $\dot{\varepsilon}$. The leftmost symbols show the data from energy-minimized simulations. Error bars indicate standard deviations from eight independent simulation runs.

Since the previous studies[21–26] demonstrated that the effects of $f$ and $\varphi_c$ on the fracture characteristics can be embedded in cycle rank $\xi$, Fig 4 examines such an argument, showing $\varepsilon_b$ and $\sigma_b$ as functions of $\xi$. As shown in the previous study, $\xi$ for the examined networks depends on $f$ and $\varphi_c$, being entirely consistent with mean-field theory[33–35] (data not shown). $\sigma_b$ is normalized by the branch point density, $\upsilon_{br} \equiv \rho/(fN_a + 1)$. Consistent with previous studies [21–26], the energy-minimized simulation data (black symbols) lie on master curves similar to power-law functions. The results from Brownian simulations exhibit similar master curves, but the curves depend on the strain rate $\dot{\varepsilon}$; $\varepsilon_b$ and $\sigma_b$ decrease with decreasing $\dot{\varepsilon}$, as discussed in Fig 3. At the lowest strain rate $\dot{\varepsilon} = 10^{-4}$ (violet), $\varepsilon_b$ and $\sigma_b$ come close to those for the energy-minimized simulations (black). The power-law exponent is independent of the strain rate, as demonstrated by panels (b) and (d), where the data are shown in double logarithmic plots.

Note that the slopes for the eye guide in panels (b) and (d) are -0.16 and 0.42, respectively, and differ from those reported earlier. For instance, in the previous report for energy-minimized simulations[26], the power-law exponents were reported as -1/2 and 1/3, respectively. Since the exponent in this study is common for energy-minimized and Brownian simulations, the difference comes from the choice of $b_c$; in the previous studies $b_c$ was chosen at $\sqrt{1.5}$, whereas this study employed $\sqrt{3.6}$ to suppress thermal degradation. The effects of this parameter on the fracture characteristics should be carefully considered but are beyond the scope of this study.

Note also that the strain rate at which the fracture characteristics closely resemble those from the energy-minimized simulations depends on the size of the system. A lower strain rate is needed for larger systems to achieve convergence to energy-minimized results. Therefore, smaller systems are better in terms of computation costs. However, due to the stochastic nature of fracture, the system size should still be sufficiently large. The system size used in this study was chosen to balance these considerations.

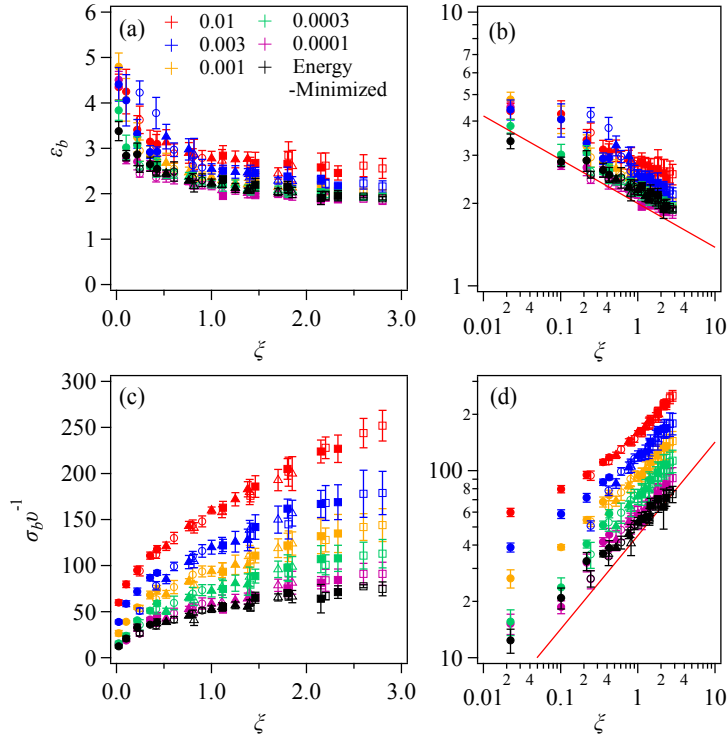

**Figure 4** Strain and stress at break, $\varepsilon_b$ (upper panels) and $\sigma_b$ (lower panels) obtained for the networks with various $f$ and $\varphi_c$ values plotted against cycle rank $\xi$.

**Conclusions**

In this study, phantom chain simulations investigated the effect of Brownian motion and strain rate on the fracture of star polymer networks. The examined networks were created from star polymer sols via end-linking reactions for various branching numbers $f$ and conversion ratios $\varphi_c$. The networks were uniaxially stretched until the break. The stress-strain curve recorded during the stretch depended on the stretch rate $\dot{\varepsilon}$, reflecting the competition between background flow and structural relaxation.

The resultant strain and stress at the break, $\varepsilon_b$ and $\sigma_b$, decreased with decreasing $\dot{\varepsilon}$. Concerning the cycle rank dependence, $\varepsilon_b$ and $\sigma_b/\nu_{br}$ (where $\nu_{br}$ is the branch point density) for various $f$ and $\varphi_c$ lie on master curves as functions of cycle rank $\xi$, demonstrating that the effects of $f$ and $\varphi_c$ on fracture characteristics are embedded in $\xi$, irrespective of the inclusion of Brownian motion and the strain rate.

Although the apparent fracture characteristics reported are similar to those for the energy-minimized case, we should note that Brownian motion has a strong influence on fracture behavior, including thermal bond breakage. We have no explanation for why cycle rank impacts various breakage processes through different kinetic pathways. We also note that mapping to experiments is yet to be investigated since several factors are missed, including excluded volume and entanglement, osmotic force, chain rigidity, etc. The physics behind the $\xi$-dependence is unknown, and the relation to other structural parameters like higher-ordered loop density[12], the minimum path length[1,10], loop opening length[36], etc., is worth discussing. Supplemental studies in such directions are ongoing, and the results will be reported elsewhere.

**Acknowledgments**
This study is partly supported by JST-CREST (JPMJCR1992) and JSPS KAKENHI (22H01189).